\documentclass[aps,prl,twocolumn,showpacs,superscriptaddress]{revtex4-1}
\usepackage{amsmath}
\usepackage{epsfig}
\usepackage{color}
\usepackage{endnotes}
\usepackage{natbib}
\usepackage[colorlinks,breaklinks]{hyperref}
\usepackage{nicefrac}
\usepackage{bm}
\usepackage{dcolumn}
\usepackage{graphics}
\usepackage[outdir=./]{epstopdf}
\usepackage{amsmath}
\usepackage{amssymb}
\usepackage{color}
\usepackage{changes}
\newcommand{\bra}{\langle}
\newcommand{\ket}{\rangle}

\newcommand{\bs}[1]{\ensuremath{\boldsymbol{#1}}}

\newcommand{\be}{\begin{equation}}
\newcommand{\ee}{\end{equation}}
\newcommand{\bea}{\begin{align}}
\newcommand{\eea}{\end{align}}
\newcommand{\beqa}{\begin{eqnarray}}
\newcommand{\eeqa}{\end{eqnarray}}

\newcount\colveccount
\newcommand*\colvec[1]{
        \global\colveccount#1
        \begin{pmatrix}
        \colvecnext
}
\def\colvecnext#1{
        #1
        \global\advance\colveccount-1
        \ifnum\colveccount>0
                \\
                \expandafter\colvecnext
        \else
                \end{pmatrix}
        \fi
}
\newcommand{\rvec}{\bs{r}}
\newcommand{\kvec}{\bs{k}}
\begin{document}

\title{The Nuclear Contact Formalism -- the Deuteron Channel}

\author{Ronen Weiss}
\affiliation{The Racah Institute of Physics, The Hebrew University, 
             Jerusalem, Israel}
\author{Nir Barnea}
\email{nir@phys.huji.ac.il}
\affiliation{The Racah Institute of Physics, The Hebrew University, 
             Jerusalem, Israel}

\date{\today}

\begin{abstract} 
The contact formalism, devised to elucidate the
important role of short-range correlations,
was recently generalized for systems with coupled channels, such as the deuteron
channel, where $s$ and $d$ waves are coupled into a $J=1$ state.
The coupling of the two channels implies two independent asymptotic
solutions and results a $2\times 2$ contact matrix.
For a strong coupling, with appropriate boundary conditions,
such as in nuclear physics, the two solutions 
degenerate into a single asymptotic wave function. 
Here, we explore the asymptotic behavior of the correlated neutron-proton pairs
in the deuteron channel using both schematic models and realistic nuclear forces.

\end{abstract}


\maketitle
{\it Introduction --}
Short range correlations (SRCs) are known to play an important role
in nuclear physics. High-momentum
tail ($k>k_F\approx 1.26$ fm$^{-1}$) originated by SRCs was identified in different nuclei 
both in theory and in experiment, see {\it e.g.}
\cite{WirSchPie14,HenSci14,AlvCioMor08,ArrHigRos12,Fomin12}, and
also \cite{Cio15, Hen_review} for reviews.
Nuclear systems are complicated systems,
composed of protons and neutrons, each with a spin degree of freedom.
In the study of SRCs in nuclear systems there are many indications
and a wide agreement in the literature
on the fact  
that, due to the tensor force,
 nuclear SRCs are dominated by proton-neutron (pn) pairs, that
can be described to a good approximation using the single $T=0$ deuteron
bound state
\cite{Piasetzky06,Subedi08,Baghdasaryan10,
Korover14,HenSci14,Schiavilla07,AlvCioMor08}.
The cross section of electron-scattering experiments,
sensitive to SRCs, is approximately
proportional to the deuteron cross-section \cite{Fomin12,ArrHigRos12,FraSar93}.
Also, the $pn$ momentum distribution, calculated numerically 
in \cite{WirSchPie14} for different nuclei, is approximately a multiplication of the
deuteron momentum distribution \cite{WeiBazBar15a,ciofi_new}.
Recently, the high momentum-tail of the one-body momentum distribution
of $A \leq 40$ nuclei was reproduced using the contact formalism \cite{WeiHen17}, where
the main contribution comes from the deuteron channel, 
using the single bound state wave function.

The contact formalism is a theoretical tool
for analyzing SRCs in quantum systems.
The contact was initially introduced 
to describe systems of two-component fermions,
obeying the zero-range condition \cite{Tan08}.
Later on, it was generalized to study nuclear systems
\cite{WeiBazBar15,WeiBazBar15a,WeiBazBar16,WeiPazBar,Hen15,BaoJun16,WeiHen17}.
Nuclear systems do not obey the zero range condition,
and as a result the contribution of all partial waves should be considered,
not only the $s$-wave contribution, and model dependent functions
must be used instead of the known zero-range two-body functions.
As a result, the nuclear contact matrices were defined and 
new relations between different nuclear quantities, which are sensitive
to nuclear SRCs, were revealed and verified
\cite{WeiBazBar15,WeiBazBar15a,WeiBazBar16,ciofi_new,WeiPazBar,WeiHen17}.
Contact matrices and contributions from different partial waves
were also considered recently to describe SRCs in other systems
\cite{HeZhou16_spectrum,ZhaHeZhou16,YosUed16}.

In \cite{WeiBar17} the contact formalism was extended 
to describe systems that, like the deuteron, are 
dominated by coupled channels.
It was found that for $n$-coupled channels, the contact
is replaced by an $n\times n$ matrix, connecting all possible asymptotic partial waves.
Therefore, instead of a universal tail to the momentum distribution, such as the $1/k^4$
tail found by Tan for zero-range interaction \cite{Tan08}, one expect a superposition
of universal terms, each having a different asymptotic behavior.

This result seems to stand in contrast with the observation that a universal deuteron-like
tail dominates nuclear SRCs. That is, the freedom suggested by the 
contact matrix of two different asymptotic functions
is some how suppressed, and we are left with a single function.
To explain this phenomena,
it was observed in \cite{WeiBar17} that by imposing a box-like boundary condition on the 
low energy spectrum, the contact matrix collapses into a single constant in the limit
of strong coupling. This means that for a strong coupling between the different channels,
the boundary condition at long distance determines the short range behavior.
For finite nuclei, such a boundary condition can be understood as the effective attraction 
induced by the surrounding nucleons on the correlated pair.

To gain more insight into the asymptotic behavior of correlated neutron-proton pairs,
we will explore here with some detail the asymptotic
behaviour of a deuteron-like system composed of $s$ and $d$-waves coupled into a $J=1$ state.
To this end, we will first review the formalism and some of the main results of 
\cite{WeiBar17}, and then present further analysis of the two-body density and two-body
momentum distribution using both ``toy model'' and realistic 
nucleon-nucleon (NN) interactions.


{\it Coupled channels contact formalism --}
Consider a nuclear system composed of $A$ nucleons.
In such a system, when two neutrons $n,n'$
approach each other, we expect 
the total wave function $\Psi(\rvec_1,\ldots,\rvec_A)$ to be dominated by 
the asympototic form
\be
  \Psi  \xrightarrow[r_{nn'}\rightarrow 0]{}\varphi(\bs{r}_{nn'})
           A(\bs{R}_{nn'},\{\bs r_k\}_{k\neq n,n'})\;.
\ee
where, 
$\bs{r}_k$ are the single-particle coordinates, 
$\bs{r}_{nn'}=\bs{r}_n-\bs{r}_{n'}$,
 $\bs{R}_{nn'}=(\bs{r}_n+\bs{r}_{n'})/2$,
$\varphi(\bs{r})$ is the universal spin-scalar $\ell=0$ zero-energy solution of the
Schr\"{o}dinger
equation of two neutrons, and $A(\bs{R}_{nn'},\{\bs r_k\}_{k\neq n,n'})$ 
is a {\it regular} function describing  
the dynamics of all other nucleons.

When considering also higher partial waves, if the different 
channels are not coupled, the asymptotic form
becomes \cite{WeiBazBar15a, HeZhou16_spectrum}
\be \label{non_coupled_asymp}
  \Psi  \xrightarrow[r_{nn'}\rightarrow 0]{}
	\sum_{\alpha} \varphi_\alpha(\bs{r}_{nn'})
           A^\alpha(\bs{R}_{nn'},\{\bs r_k\}_{k\neq n,n'})\;.
\ee
The sum over $\alpha$ indicates the different channels,
which in the case of two nucleons are given by 
 $\alpha=(\ell,s)j\,m$.
When considering coupled channels, the quantum numbers
indicated by $\alpha$  are not necessarily sufficient
to describe the universal
part of the asymptotic wave-function.

To illuminate the difference between the coupled and uncoupled cases
we shall focus on the deuteron channel.
Generalization of the following arguments 
to any other channels is strait forward.
For a neutron-proton pair,
residing in a deuteron-like state, the wave function is composed
of an $s$ and $d$ channels,
given by
  $|s\ket=|(\ell=0,s=1)j=1\,m\ket$ and
  $|d\ket=|(\ell=2,s=1)j=1\,m\ket$,
and the general solution to the Schr\"{o}dinger equation is
a superposition of two independent solutions,
\be \label{coupled_sol}
\varphi^{a}(\bs{r}) = \varphi^{a}_{s}(r) |s \ket +
 		          \varphi^{a}_{d}(r) |d \ket.
\ee
The indices $a=1,2$ stands for the two independent solutions.
Each of these solutions is a mixture of both channels,
$|s \ket$ and $|d \ket$. 
Now it is evident that when a neutron $n$ and a proton $p$ 
approach each other, the asymptotic wave function, dominated by
the deuteron channel, will take the form
\be \label{coupled_asymp}
\Psi \xrightarrow[r_{np}\rightarrow 0]{}
	\sum_{a=1,2} \varphi^a(\bs{r}_{np})
           A^a(\bs{R}_{np},\{\bs r_k\}_{k\neq n,p})\;.
\ee
Notice that it is different than the asymptotic form 
given in Eq. \eqref{non_coupled_asymp}, since Eq.
\eqref{coupled_asymp} includes two different functions
$\varphi^1_s(r)$ and $\varphi^2_s(r)$,
and each of them is generally coupled to a different $A^a$
function.
Since the nuclear wave function $\Psi$ has a well-defined total
angular momentum $J$, a summation over $m$ is also required here
(see Ref. \cite{WeiBazBar15a}). We omit it here for simplicity, since
it does not affect our conclusions.

The above asymptotic form leads to the definition 
of a $2 \times 2$ contact matrix for the case of two coupled channels:
\be \label{coupled_contacts}
C^{ab}=16\pi^2 NZ \bra A^a | A^b \ket,
\ee
where $a,b=1,2$. Notice that $A^1$ and $A^2$ are generally 
not orthogonal. 
Previous  studies, that have defined a matrix of contacts
\cite{WeiBazBar15a,ZhaHeZhou16,YosUed16},
implicitly assumed that asymptotically the potential does not couple different channels.

As presented in \cite{WeiBar17}, this asymptotic form and contact matrix can
be used to derive different contact relations.
We review the deuteron-channel contribution to the momentum and density distributions.
The single-particle momentum distribution, $n(k)=\int d\hat{k} n(\kvec)$,
describing the probability to find a particle with momentum $k$, is given 
asymptotically by
\be \label{1b_mom}
n(k) \xrightarrow[k \rightarrow \infty]{}
\sum_{a,b=1,2} \left( \tilde{\varphi}^{a*}_s(k)\tilde{\varphi}^{b}_s(k)
+  \tilde{\varphi}^{a*}_d(k)\tilde{\varphi}^{b}_d(k)
 \right) \frac{C^{ab}}{16\pi^2},
\ee
where 
$n(\kvec)$ is normalized to the number of nucleons in the system $A$, and
$\tilde{\varphi}^a_\alpha(k)$
is the Fourier transform of $\varphi^a_\alpha(r)$.
The two-particle momentum distribution
$F(k)=\int d\hat{k} F(\kvec)$, which describes the probability to find an 
$np$
particle pair with relative momentum $k$, 
is given by
\be \label{2b_mom}
  F(k) \xrightarrow[k \rightarrow \infty]{}
    \sum_{a,b=1,2} \left( \tilde{\varphi}^{a*}_s(k)\tilde{\varphi}^{b}_s(k)
   + \tilde{\varphi}^{a*}_d(k)\tilde{\varphi}^{b}_d(k)
                 \right) \frac{C^{ab}}{16\pi^2}.
\ee
Finally, the asymptotic probability to find a pair of $pn$ particles with relative distance 
$r$, $\rho(r)=\int d\hat{r} \rho(\rvec)$, is given by
\be \label{density}
\rho(r) \xrightarrow[r \rightarrow 0]{}
\sum_{a,b=1,2} \left( \varphi^{a*}_s(r) \varphi^{b}_s(r)
+  \varphi^{a*}_d(r) \varphi^{b}_d(r)
 \right) \frac{C^{ab}}{16\pi^2}.
\ee
$\rho(r)$ and $F(k)$ are normalized to the number of pairs.

From these results it is evident that if the two solution 
$\varphi^a$, $a=1,2$, have different asymptotic forms,
then, depending on the explicit form of the contact matrix \eqref{coupled_contacts},
each nucleus might have a different asymptotic momentum or density distributions.
For example, if we will
compare two different eigenstates, $\Psi_1$ and $\Psi_2$,
and look on the ratio of the two corresponding momentum distributions,
$n_1(k)/n_2(k)$, this ratio will generally not obtain a constant value
for high momentum. This is because the values of the different
four contacts $C^{ab}$ can be different for each state, and the
$k$-dependence will not generally disappear, same 
for $F(k)$ and $\rho(r)$.
As mentioned before, this is the result that seems to contradict the universal
deuteron-like behavior of nuclear SRCs. Since nuclear momentum and density
distributions are reproduced using the single bound-state deuteron wave function,
such a ratio of two distributions will have an asymptotic constant behavior,
in contrast to the discussion above.

In a  mean field picture, the nucleons in the atomic nucleus are subject to an
attractive potential whose magnitude is of the order of 
$\epsilon_F\approx 30 \: \rm{MeV}$. Consequently, the universal wave function $\varphi$,
Eq. \eqref{coupled_sol}, is also subject to such an attractive potential, and all bound
nucleon pairs will have an exponentially-decaying long-range tail. 
This effect of the mean field potential was approximated in \cite{WeiBar17}
through hard wall boundary condition. There, it was found that if the coupling
between the $s$ and $d$ channels is strong enough, as is the case for nuclear-physics,
 all the low-laying two-body states 
with energy below about 30 MeV have the same asymptotic form.

Consequently, considering only the deuteron-channel contribution,
in finite nuclei it is expected that
\be \label{1b_mom_D}
   n(k) \xrightarrow[k \rightarrow \infty]{}
            \left(\tilde{\varphi}^{*D}_s(k)\tilde{\varphi}^{D}_s(k)
                 +\tilde{\varphi}^{*D}_d(k)\tilde{\varphi}^{D}_d(k)\right)
            \frac{C}{16\pi^2},
\ee
\be \label{2b_mom_D}
  F(k) \xrightarrow[k \rightarrow \infty]{}
            \left(\tilde{\varphi}^{*D}_s(k)\tilde{\varphi}^{D}_s(k)
                 +\tilde{\varphi}^{*D}_d(k)\tilde{\varphi}^{D}_d(k)\right)
              \frac{C}{16\pi^2},
\ee
and
\be \label{density}
  \rho(r) \xrightarrow[r \rightarrow 0]{}
            \left({\varphi}^{*D}_s(r){\varphi}^{D}_s(r)
                 +{\varphi}^{*D}_d(r){\varphi}^{D}_d(r)\right)
          \frac{C}{16\pi^2},
\ee
where $\varphi^{D}_s(r),\varphi^{D}_d(r)$ are the $s,d$ components of the deuotron 
wave function, and $\tilde\varphi^{D}_s(k),\tilde\varphi^{D}_d(k)$ are their Fourier
transform.
This result solves the contradiction between the known nature of
nuclear SRCs and the predictions of the coupled-channels contact theory.

{\it The asymptotic wave-function -- }
To solve the Schr\"{o}dinger equation for the coupled $s,d$-channels
we can integrate the radial equation from $r=0$ outwards starting 
with one of the two distinct boundary conditions
\begin{align}
 \frac{d(r\varphi_s^1)}{dr}& =1  &\frac{d^3(r\varphi_d^1)}{dr^3}&=0 &\cr
 \frac{d(r\varphi_s^2)}{dr}& =0  &\frac{d^3(r\varphi_d^2)}{dr^3}&=1 \;.&
\end{align}
For a scattering, positive energy, case there are no further conditions on the 
wave-function, and therefore $\varphi^1, \varphi^2$ or any linear combination
\be
  \varphi(\bs r) = c_1 \colvec{2}{\varphi^1_s(r)}{\varphi^1_d(r)}
                 + c_2 \colvec{2}{\varphi^2_s(r)}{\varphi^2_d(r)},
\ee
with $c_1$ and $c_2$ being free coefficients,
are legitimate physical solutions. 
Adding a bound-state 
boundary condition of a vanishing wave-function at $r \rightarrow \infty$ leads
to energy quantization,
and for each allowed energy
there is only one solution characterized by the value of the ratio
$\eta \equiv c_2/c_1$.

\begin{figure}\begin{center}
\includegraphics[width=8.6 cm]{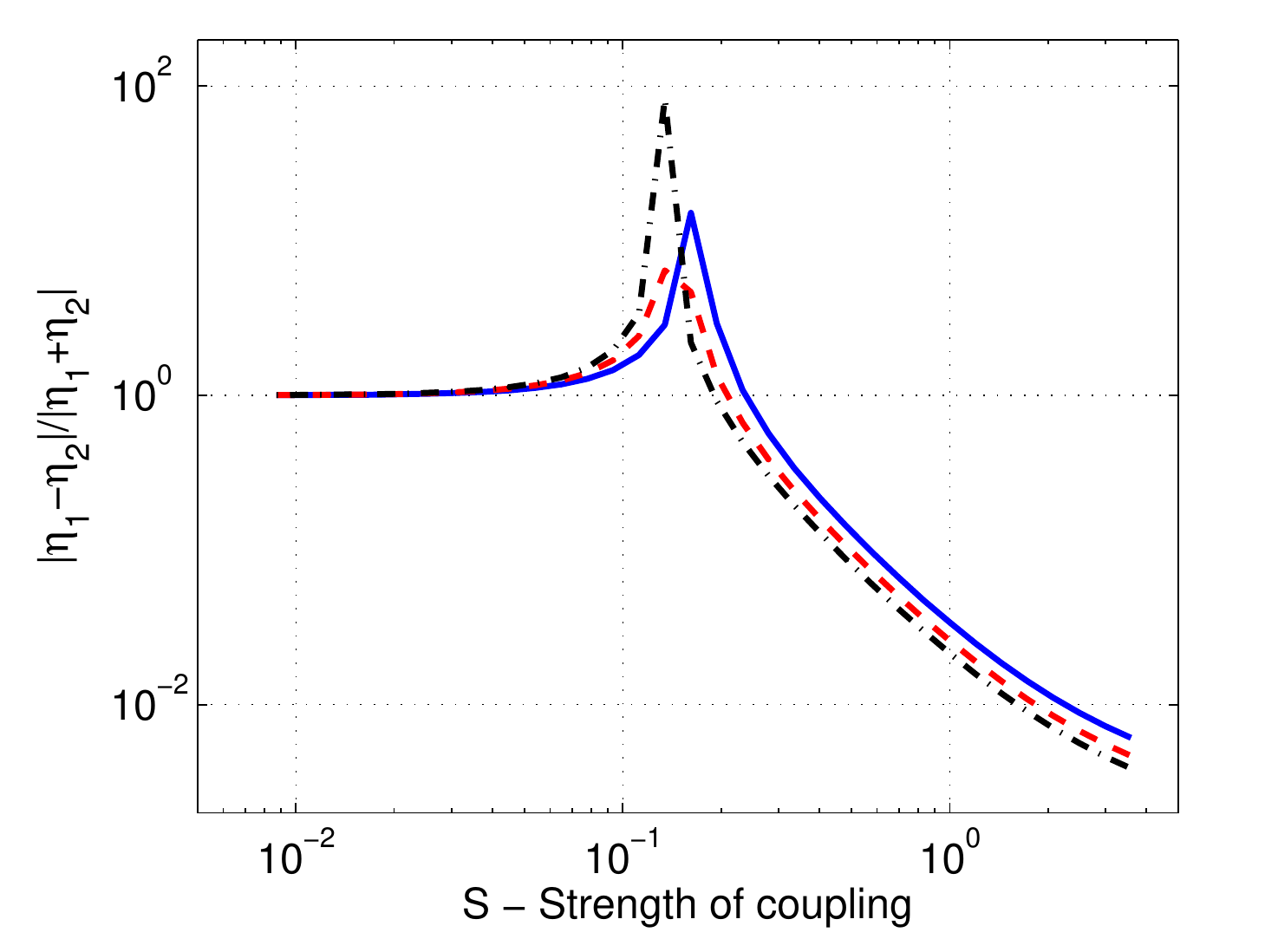} 
\caption{\label{strngth} 
Comparison between the values of 
 $\eta(E) \equiv c_2(E)/c_1(E)$ for the first
two allowed energies in a box with radius $R$
as a function of $S$, using
the simple ``toy model' potential.
We used $\Lambda^2=3$ fm$^{-2}$ and the scattering
length was kept constant $a_s=10$ fm.
The blue, red and black lines
are for $R=15, 25, 35$ fm.
}
\end{center}\end{figure}

In order to see the implications of such a boundary condition,
we will first use the simple ``toy model'' composed of a simple gaussian potential with 
$s,d$ coupling introduced in \cite{WeiBar17},
\begin{align} \label{GaussP}
  V_{s,s}=V_{d,d} & =-V_0 \exp(-\Lambda^2 r^2) & \cr
  V_{s,d}=V_{d,s} & =-S V_0 \exp(-\Lambda^2 r^2) &
\end{align}
and solve the Schr\"{o}dinger equation for this potential
in a spherical hard wall box of radius $R$.
For a given strength $S$, $V_0$ is tuned to produce
a constant scattering length, 
and the ratio of $\eta_i$ for the $i^{th}$ energy level
is extracted from the corresponding wave function.
The results for the ratio 
$|\eta_1(S)-\eta_2(S)|/|\eta_1(S)+\eta_2(S)|$ 
comparing the $\eta$ values of the first two energy levels are presented 
in Fig. \ref{strngth}.
The calculations were done for scattering length $a_s=10\:\rm{fm}$
and different values of $R$.
In the non-coupling limit $S\longrightarrow 0$, the lowest energy state
is a pure $s$-wave, $\eta \rightarrow 0$,  and the $2^{nd}$ excited state
is a pure $d$-wave, $\eta \rightarrow \infty$. Consequently the ratio
$|\eta_1(S)-\eta_2(S)|/|\eta_1(S)+\eta_2(S)|\longrightarrow 1$, as can be seen in the figure.
On the other hand, as was already observed in \cite{WeiBar17}, the ratio
$|\eta_1(S)-\eta_2(S)|/|\eta_1(S)+\eta_2(S)|\longrightarrow 0$ as the coupling $S$ becomes 
stronger and stronger, which means that the short-range behavior
of the two wavefunctions become identical 
in the strong-coupling limit ($\eta_1 \approx \eta_2$).
It should be emphasized that this phenomenon is due to the boundary condition at $R$,
and does not depend on the exact value of $R$.
For scattering states there is always a complete freedom to choose the parameters $c_1$
and $c_2$ at will.

\begin{figure}\begin{center}
\includegraphics[width=8.6 cm]{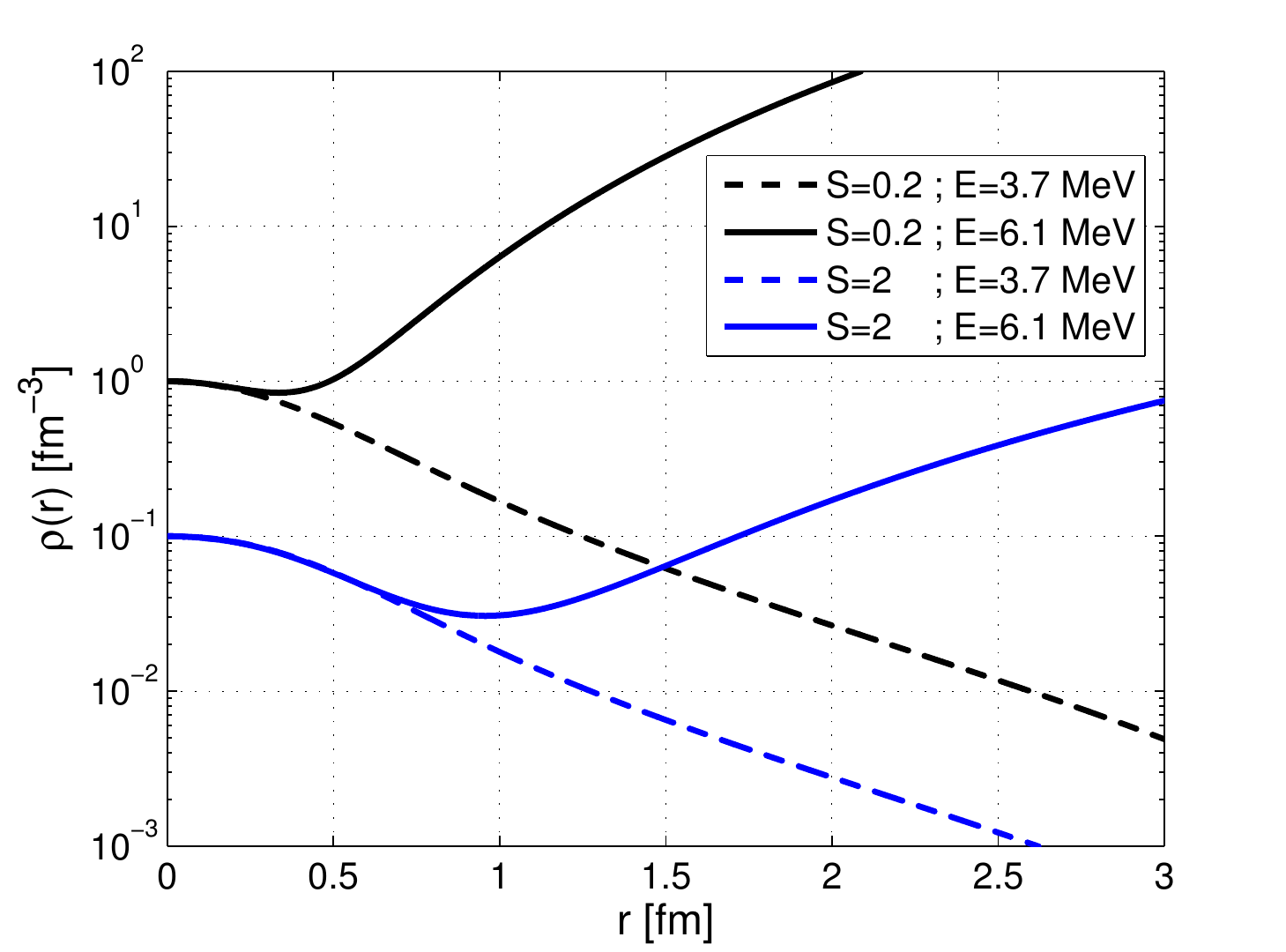}
\caption{\label{rho_S} 
The densities of the two lowest energy states in the ``toy model''
with a box boundary condition at $R=15$ fm, and $S=0.2$ or $S=2$.
The scattering length was fixed to
$a_s=10$ fm, and $\Lambda^2=3$ fm$^{-2}$.
}
\end{center}\end{figure}

We now turn to study the effect of the coupling strength on the density.
In Fig. \ref{rho_S} we present the densities of the two lowest energy states
obeying box boundary condition, with two different coupling values $S=0.2$ and $S=2$.
The calculation was carried out for a fixed scattering length $a_s=10$ fm.
In this figure it can be seen that 
for the strong coupling ($S=2$), $\rho(r)$ is similar for these two energy states
up to almost $r=0.8$ fm. On the other hand, for $S=0.2$, the solutions corresponding 
to the two states
behave differently starting from a smaller distance, around $r\approx 0.3$ fm.
The similar behavior for smaller distances is because in such small distances
only the $s$-wave component of the solutions is significant, while the $d$-wave
component goes to zero in the origin.
We therefore deduce that also in an $A$-body system with intermediate values
of coupling strength,
when two particles get close to each other, they are not restricted
to behave  like the bound "deuteron", but rather as any of the two solutions. As the coupling
becomes stronger, a convergence towards unique short range behavior is again observed. 

\begin{figure}\begin{center}
\includegraphics[width=8.6 cm]{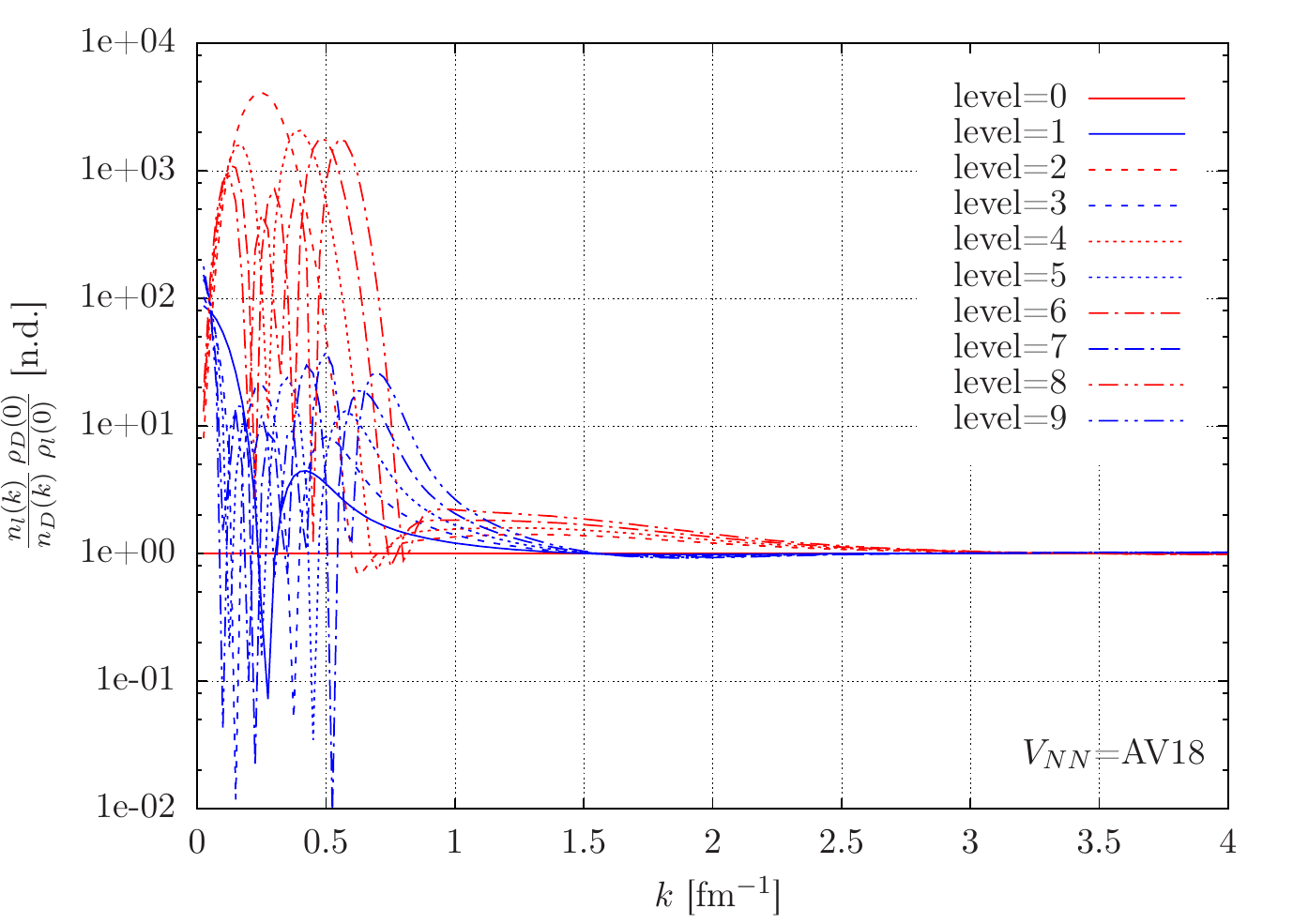}
\caption{\label{mmnt_av18_ho} Momentum distribution ratios for the first few two-body levels
in an external HO potential. Here we have used the AV18 NN potential.
}
\end{center}\end{figure}

\begin{figure}\begin{center}
\includegraphics[width=8.6 cm]{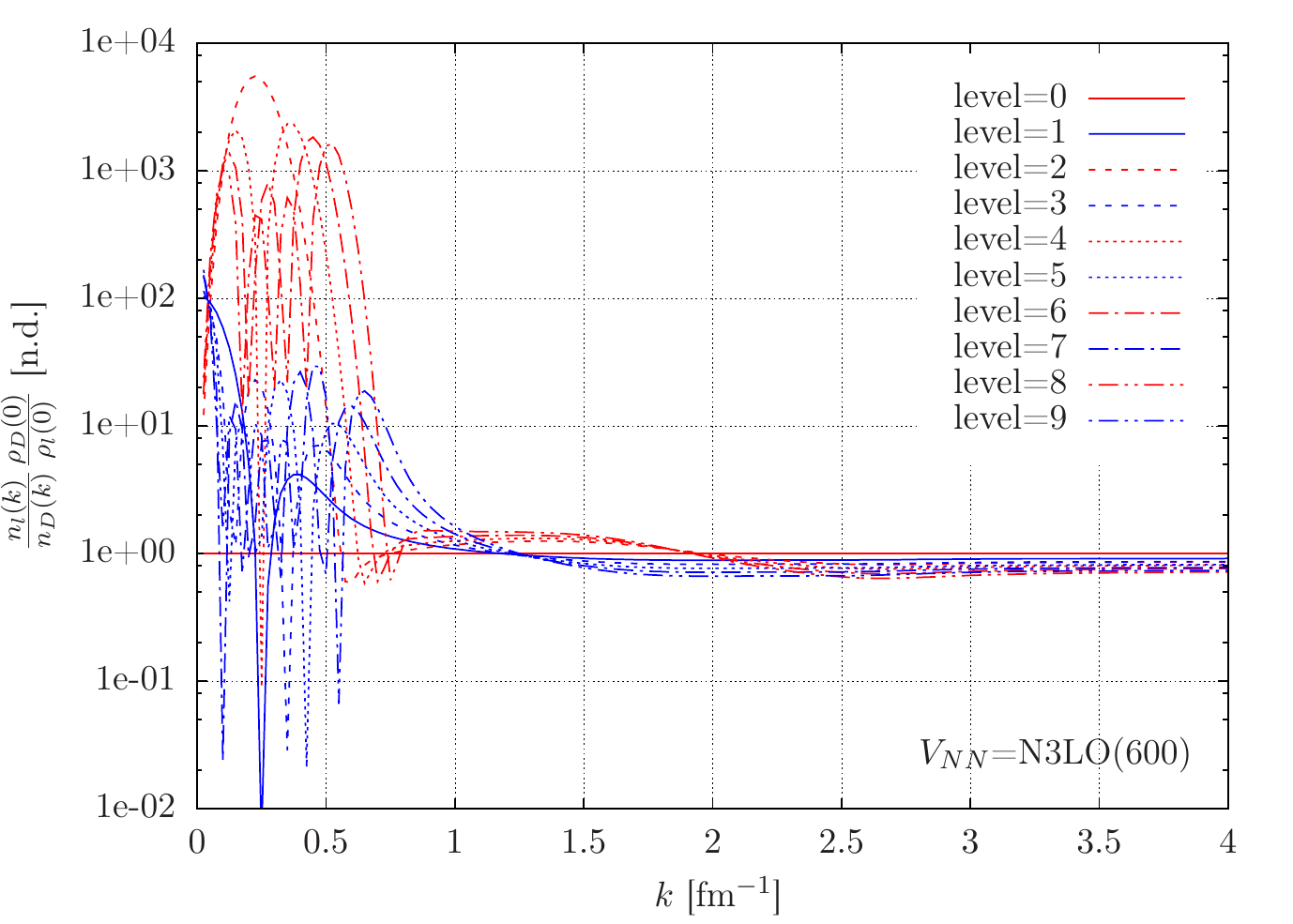}
\caption{\label{mmnt_n3lo_ho} Momentum distribution ratios for the first few two-body levels
in an external HO potential. Here we have used the chiral EFT N3LO(600) NN potential.
}
\end{center}\end{figure}

Now, let us turn and  study the implications of the boundary condition
on the neutron-proton momentum distribution using 
the realistic NN potential model AV18 \cite{av18} and the chiral EFT
force N3LO(600), see \cite{N3LO} and ref. therein.
To this end, we have solved the Schr\"{o}dinger equation
in an external harmonic oscillator (HO) potential, that simulates the effective
mean field potential the nucleons fill inside the nucleus.
Similar to the box boundary condition,
this external potential also leads to the quantization of
the continuum solutions. 
In figures \ref{mmnt_av18_ho} for AV18 and \ref{mmnt_n3lo_ho}
for N3LO(600), 
we present the normalized momentum distributions ratios 
$$\left(\frac{n_l(k)}{n_D(k)}\right)\left( \frac{\rho_D(0)}{\rho_l(0)}\right)$$
of the first 10 levels (subscript $l$)
compared to the deuteron solution (subscript $D$), where $\rho$ is the coordinate-space density. 
The levels spread the energy range $-2.2 \;\rm{MeV} \leq E_l \leq 20 \;\rm{MeV}$, due to 
the value of $\hbar\omega_{HO}=0.1\;\rm{MeV}$.
One can see that also here, due to the external potential,
all the low energy continuum solutions are similar to the bound deuteron solution for 
large momenta. 
We note that the two different nucleon-nucleon potentials used in these calculations
lead to the same conclusion.

Inspecting the figures, one can observe the grouping of the levels into two color
groups having slightly different behavior at the momentum range 
$1.5\;\rm{fm}^{-1} \leq k \leq 2.5\;\rm{fm}^{-1}$. 
This grouping can be attributed to the different
$\eta$ values. For $k\geq 2.5\;\rm{fm}^{-1}$ all the solutions collapse into a single 
solution proportional to the deuteron.
One might argue that what we see is the $s$-wave dominance, however, it should be noted
that at 
$k \approx 4\;\rm{fm}^{-1}$ the
$s$-wave and $d$-wave contributions to $n(k)$ are roughly equal.


{\it Summary -- } 
Summing up, we have seen here and in \cite{WeiBar17}, that 
without imposing a boundary condition at $r \rightarrow \infty$,
the contact formalism for two coupled channels requires a $2 \times 2$
contact matrix and two asymptotic functions. This seems to contradict
 the known features of nuclear SRCs that are dominated by the deuteron 
channel with the single bound-state deuteron wavefunction.
Adding such a boundary condition resolve this tension,
because, in the strong-coupling limit, it results in a collapse of the
contact matrix to a single contact and a single asymptotic function.
This boundary condition can be interpreted as the effective mean-field
potential applied on the correlated pair due to the remaining particles in the nucleus.
In the case of a weak coupling, two asymptotic functions are still generally needed.

Using a ``toy model'' and a hard-wall boundary condition,
it was shown that for a strong coupling term in the potential,
the densities of the two lowest energy states coincide over a significant
range. In addition, using two different realistic nuclear forces, and in the presence
of an external HO potential (instead of the hard wall),
the asymptotic momentum distribution of the first
few positive energy solutions coincide with the bound-state deuteron high-momentum
tail. This indicates that the collapse to a single asymptotic wave function
in the strong-coupling limit is a general
phenomenon, independent of the exact NN potential or boundary condition.

\begin{acknowledgments}
This work was supported by the Pazy Foundation.
\end{acknowledgments}


\end{document}